\documentclass[sigconf]{acmart}
\renewcommand\footnotetextcopyrightpermission[1]{} 
\usepackage{subcaption}
\usepackage{todonotes}
\usepackage[skins]{tcolorbox}
\usepackage{listings}
\usepackage{xcolor}
\usepackage[utf8]{inputenc}
\usepackage{tikz}
\usetikzlibrary{decorations.pathreplacing,calc,shapes,positioning,tikzmark}

\newsavebox{\tempbox}

\AtBeginDocument{%
  \providecommand\BibTeX{{%
    \normalfont B\kern-0.5em{\scshape i\kern-0.25em b}\kern-0.8em\TeX}}}

\acmConference[KLEE 2022]{3rd International KLEE Workshop on Symbolic Execution}{Sept 15--16,
  2022}{London, UK}
\begin{document}
\title{Presentation: SymDefFix - Sound Automatic Repair Using Symbolic Execution}

\author{Tareq Mohammed Nazir and Martin Pinzger}
\orcid{}
\affiliation{%
  \institution{Universit\"at Klagenfurt}
  \city{Klagenfurt}
  \state{Carinthia}
  \country{Austria}
}
\email{{tareq.nazir,martin.pinzger}@aau.at}
\settopmatter{printacmref=false}



\begin{abstract}
In this presentation, we introduce our constraint-based repair approach, 
called SymDefFix. SymDefFix is based on ExtractFix \cite{gao2021beyond} 
and replaces the dynamic analysis steps of ExtractFix to detect the error
and find the potential fix locations in an input program with symbolic execution. 
We first briefly motivate and introduce our 
modifications of ExtractFix, and then demonstrate
it with an example. 
\end{abstract}

\maketitle


\section{Introduction}

Fixing software bugs and vulnerabilities in programs is a time consuming task 
that researchers try to automate. During the last decade they proposed 
several approaches to address this issue. Most of them are based on 
executing a test suite to detect a bug or vulnerability and synthesize a 
patch for it \cite{le2011genprog, qi2014strength, le2017jfix,mechtaev2018test, long2016automatic, nguyen2013semfix, xuan2016nopol, mechtaev2015directfix, mechtaev2016angelix, le2017jfix}. 
These approaches suffer from the problem of 
generating low-quality patches that over-fit the given test suite. 
ExtractFix \cite{gao2021beyond} is an approach that addresses the over-fitting problem via 
symbolic reasoning. It uses sanitizers to extract the underlying cause of 
a vulnerability and uses that information to first find potential fix
locations and then to generate a patch at each location. However, ExtractFix 
relies on the information from executing one test case that triggers the 
bug. In a first manual evaluation, we found that this limits the 
capabilities of ExtractFix and led us to the following research
hypothesis: 
\definecolor{light-gray}{gray}{0.95}
\begin{tcolorbox}[skin=widget,
boxrule=0.5mm,
colframe=black,
colback=light-gray,
width=(1.0\linewidth),before=\hfill,after=\hfill,before
skip=10pt plus 1pt,after skip=10pt plus 1pt]
We can use symbolic execution to obtain more complete information \t
about the bug and use that information to find more accurate \t 
fix locations and generate more accurate patches.
\end{tcolorbox}

\begin{figure*}[btp]
    \centering
        \includegraphics[width=0.68\textwidth]{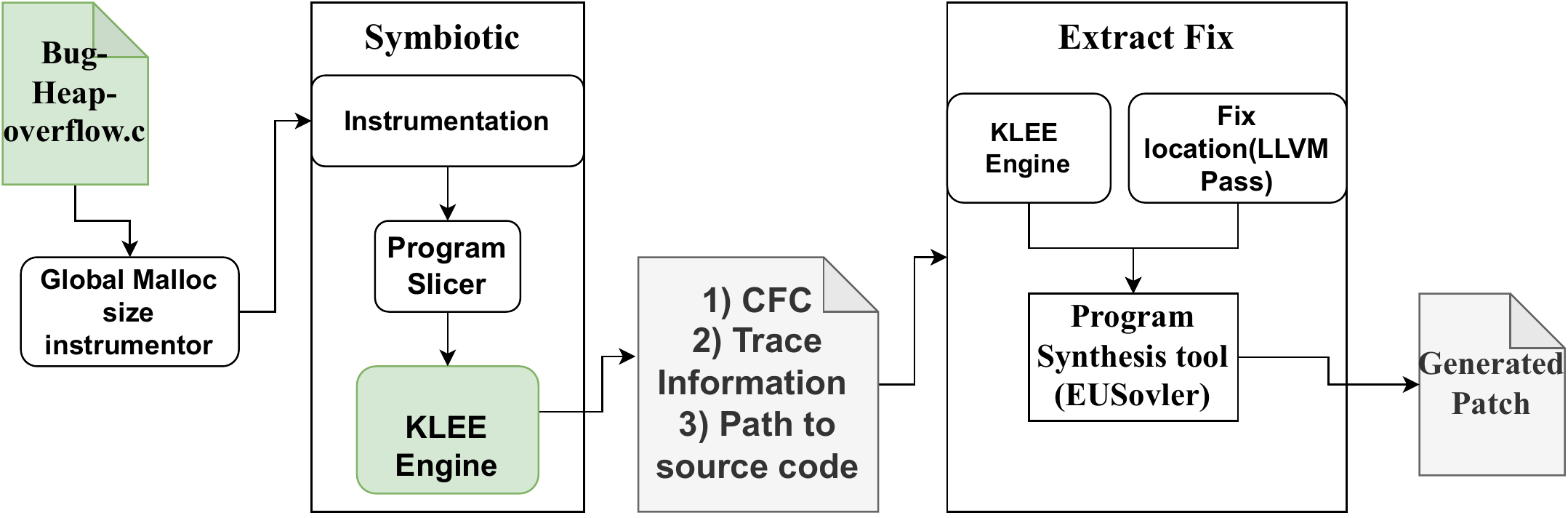}
        \caption{Overview of the SymDefFix approach.}
        \label{fig:workflow1}
\end{figure*}

In the following sections, we briefly present our modifications
of ExtractFix and showcase our approach, that we call SymDefFix, 
with an example.

\section{Symbolically Defined Fix (SymDefFix)}
ExtractFix reads a C program, instruments it with a sanitizer, and then executes
the given test case to generate a crash. When a crash occurs, the sanitizer outputs
the crash location and the constraint that was found by the sanitizer to be 
violated, called
the crash-free constraint (CFC). In addition, it records a call trace to the function
in which the crash occurs. 
Next, using the CFC and the trace recorded for the failing test case,
it identifies potential fix locations using control and data flow 
analysis. Then, it propagates the CFC to each fix location and, finally,  
it generates a patch that complies with the CFC at each fix location.
For more information on ExtractFix, we refer the reader to \cite{gao2021beyond}. 

We tried ExtractFix on several examples, provided with the ExtractFix tool, and
adapted from the SV-COMP\footnote{\url{https://sv-comp.sosy-lab.org/2021/results/results-verified/}} 
benchmark with a focus on samples from the memsafety category. 
Through that, we found two main limitations of ExtractFix: first, 
developers need to provide a test case that triggers the error; second, only the trace
that leads to the error is considered for finding the fix locations. The latter 
ignores cases in which multiple traces (i.e paths) might lead to the error.

In our approach, called SymDefFix, we address both shortcomings by using a static formal 
verification tool to detect the error(s) considering all potential program paths. 
%
%
Figure \ref{fig:workflow1} presents our approach highlighting our modifications of
the original ExtractFix. In particular, we replace ExtractFix's step to extract the CFC  
and the execution trace 
with Symbiotic \cite{chalupa2016symbiotic}. Symbiotic is a framework for static program 
analysis and verification that uses KLEE.
As shown on the left side of Figure \ref{fig:workflow1}
Symbiotic first instruments and slices the code and then it symbolically 
executes the remaining 
code using KLEE to detect any errors that exist in the code. The
usage of KLEE as part of Symbiotic is illustrated in detail 
in \cite{chalupa2016symbiotic}. 

We modified the KLEE implementation inside Symbiotic to output three types of information: 
a) crash free constraints (CFC); b) trace information, and c) path to the source file instrumented by Symbiotic. 
The CFC and the trace information are the same as output by ExtractFix, 
but this time determined using symbolic execution. In contrast to ExtractFix, our approach
explores \emph{all} possible paths and consequently finds all paths that can lead to the error plus 
the corresponding CFC. All three outputs are then provided to the original ExtractFix 
to find the candidate fix locations 
and the weak preconditions. These two inputs then are provided to program synthesis tool 
EUSolver \cite{alur2017scaling} that, for each fix location, generates a valid patch 
that satisfies the CFC under all inputs.

\section{Example}
We demonstrate the feasibility of our approach with the heap-overflow example shown in 
Figure \ref{fig:heap-overflow}.
Note, instead of reading the command line argument, which is used by ExtractFix's 
test case, we hard coded the 
size of \texttt{content} variable to be 10. 
Furthermore note, in the current version of SymDefFix, we are using ExtractFix's 
Global Malloc size instrumentor (GSInserter) that  
introduces a global variable to represent the size of the memory allocated with \texttt{malloc}.Analyzing this example with SymDefFix, Symbiotic correctly detects an error at line 19 and outputs:

\begin{figure}[btp]
        \includegraphics[width=0.47\textwidth]{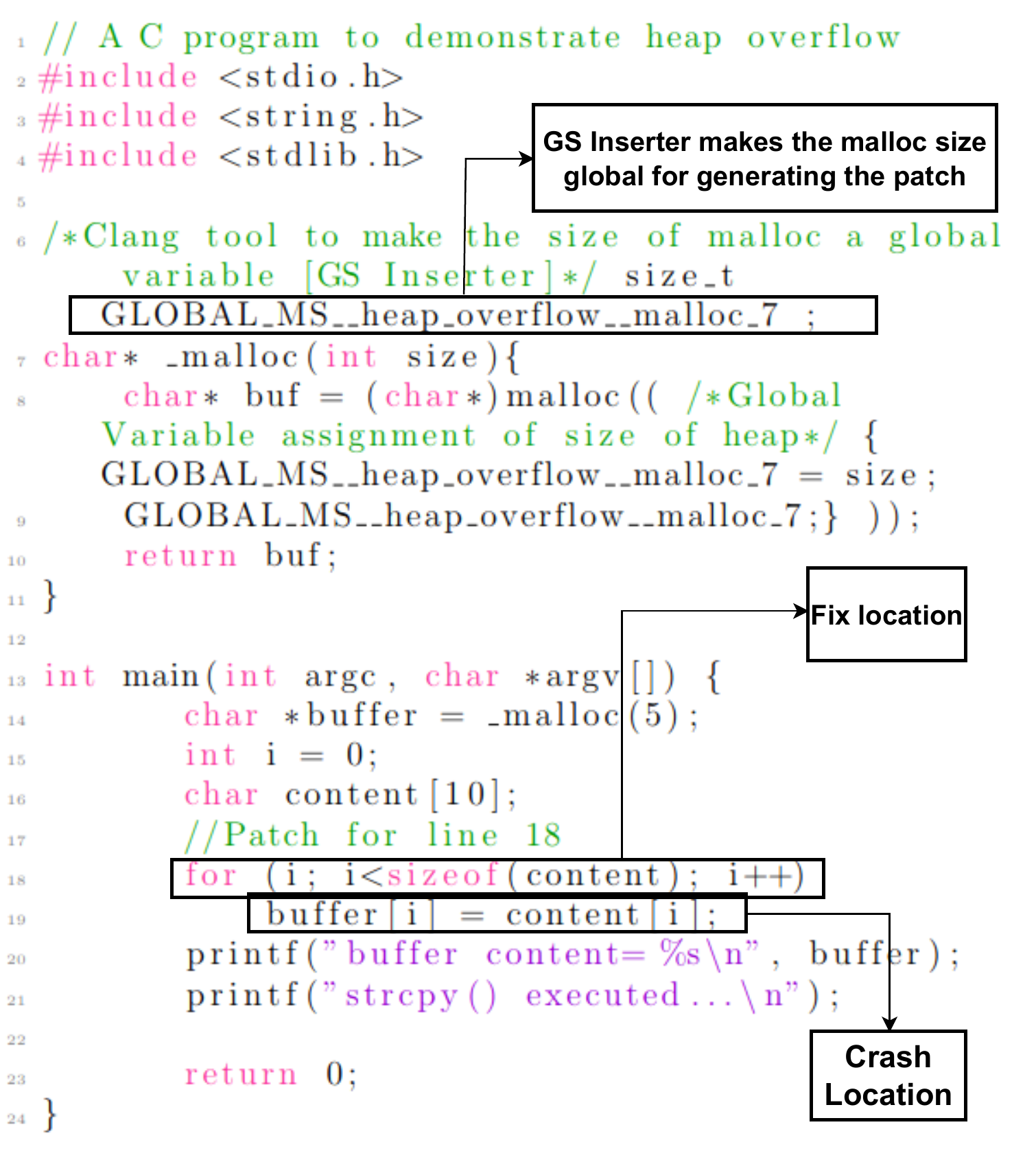}
        \caption{Heap-Overflow example highlighting the crash and candidate fix location.}
        \label{fig:heap-overflow}
\end{figure}


%
\begin{itemize}
\item[CFC:] $access(buffer) < base(buffer)+size(buffer)$
\item[Trace:] ["IN, "main"]
\item[Path:] ../tmp/output.txt
\end{itemize}

The example violates the CFC in line 19 when it assigns the sixth element of \texttt{content} to 
\texttt{buffer}, because the size of the \texttt{buffer} is only five bytes.
%
This information is then provided to ExtractFix that determines the potential 
fix locations and uses the EUSolver\cite{alur2017scaling} to generate the patch(es). 
The patch for this 
example is presented in Listing \ref{code:patch}. It modifies the condition in the 
for loop at line 19 adding the boundary condition of the heap size that is stored in the variable "GLOBAL\_MS\_\_heap\_overflow\_\_malloc\_7". 

\begin{lstlisting}[basicstyle=\small,language=C++, caption= Generated Patch,label=code:patch]
- for (i; i<sizeof(content); i++)
---
+ for (i;(((i)<sizeof(content))&&((i)<
(GLOBAL_MS__heap_overflow__malloc_7)));
i++)
\end{lstlisting}


\section{Future Direction and Goals}
Currently, we only consider the information from one symbolically executed path to 
determine the fix locations and generate patches. As a next step, we want to 
consider all paths explored by KLEE (inside Symbiotic) for this step to further investigate our
hypothesis stated in the introduction. Furthermore, we will extend our approach to consider more error types, such as divide-by-zero.

\bibliographystyle{ACM-Reference-Format}
\bibliography{main-ref}
\appendix

\end{document}